**Sergey Yekimov**
Department of Trade and Finance, Faculty of Economics and Management, Czech University of Life Sciences Prague, Kamycka 129, 16500, Praha - Suchdol, Czech Republic

# Interpolation of numerical series by the Fermat–Torricelli point construction method on the example of the numerical series of inflation in the Czech Republic in 2011-2021

## Abstract

The use of regression analysis for processing experimental data is fraught with certain difficulties, which, when models are constructed, are associated with assumptions, and there is a normal law of error distribution and variables are statistically independent.

In practice , these conditions do not always take place .  This may cause the constructed economic and mathematical model to have no practical value.  As an alternative approach to the study of numerical series, according to the author, smoothing of numerical series using Fermat-Torricelli points with subsequent interpolation of these points by series of exponents could be used.

The use of exponential series for interpolating numerical series makes it possible to achieve the accuracy of model construction no worse than regression analysis . At the same time, the interpolation by series of exponents does not require the statistical material that the errors of the numerical series obey the normal distribution law, and statistical independence of variables is also not required. Interpolation of numerical series by exponential series represents a "black box" type model, that is, only input parameters and output parameters matter.

## Keywords

## Introduction

In the process of carrying out mathematical calculations, there is often a need for differentiation and integration of functions, this is one of the reasons that requires the transformation of functions given in tabular form to the form of an analytical function. In scientific research, the researcher is required to predict the course of certain processes in the future. By approximating the numerical series with an empirical formula, it becomes possible to extend current and past trends to future processes.

According to [1], the approximation should assume that the data calculated using the formula should, with the required accuracy, correspond to the data presented in the table.

According to [2], achieving the necessary accuracy at the nodal points does not guarantee that in the interval between the nodal points, the values calculated on the basis of the formula will correspond to the actual values.

According to [3], the model used for interpolation should have its statistical characteristics analyzed, as well as the correspondence of the properties of the points located between the nodal points to the observed values.

According to [4], if the regression curve has a very complex form, then this may lead to the fact that this curve may be devoid of practical meaning.

According to [5], when choosing the criterion for the optimal approximation of a function, it is possible to use as a criterion for the optimality of the approximation the exact coincidence of the values of the approximating fuction with its tabular values in some nodes. As approximating functions, it is advisable to use linear combinations of some power functions, trigonometric functions sine and cosine, as well as exponential functions.

According to [6], the least squares method has become widely used for approximating tabular data.

It was first used in the nineteenth century by Legendre and Gauss.

The mathematical justification of the applicability of the least squares method was made by Markov [7] and Kolmogorov [8]. The use of the least squares method is justified if:

1. Observation errors obey the normal distribution.
2. With each observation, the mathematical expectation of a random error is zero.
3. With each observation, the amount of variance is unchanged.
4. Errors in any two observations are independent of each other.

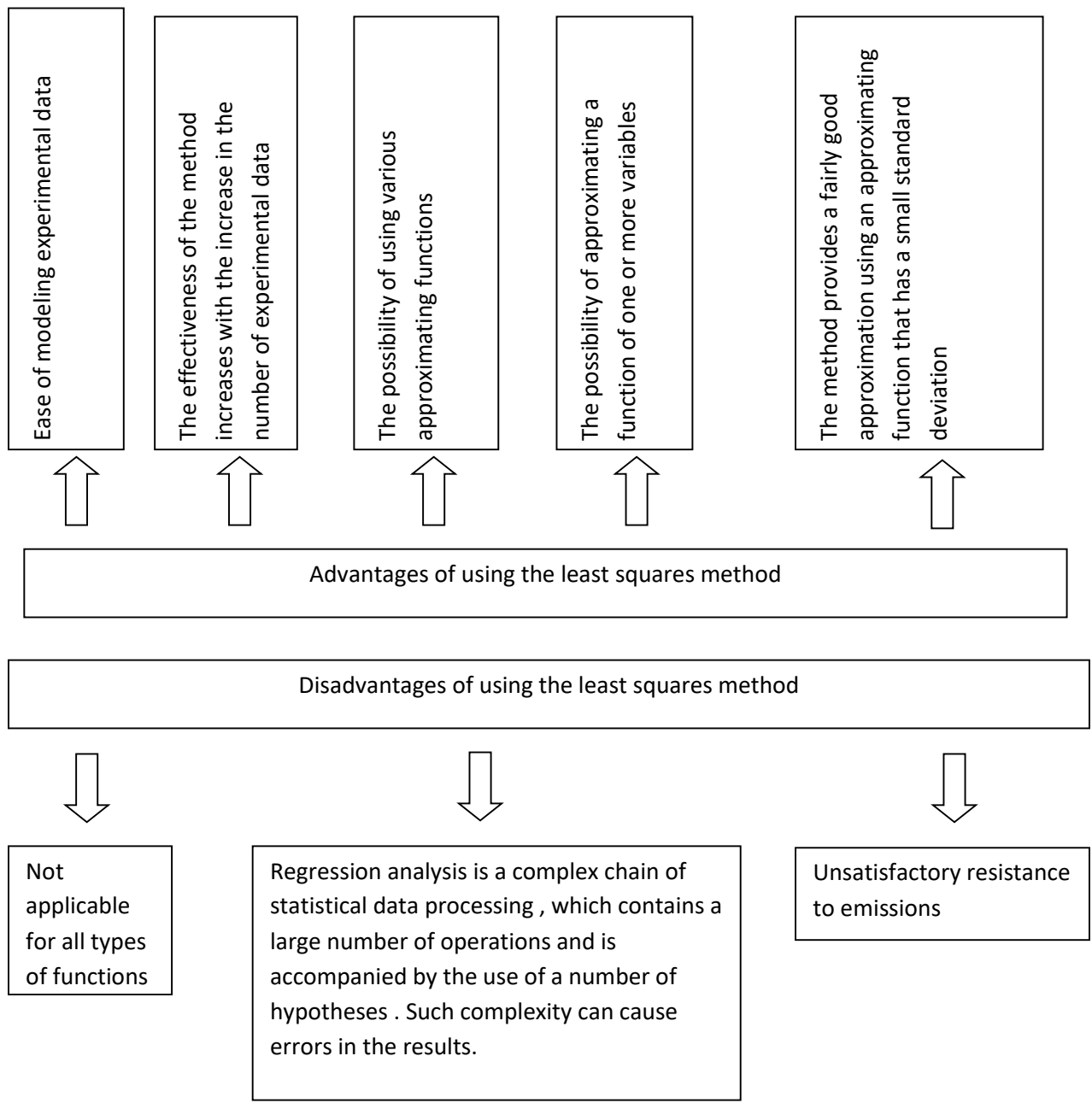

Fig. 1. Advantages and disadvantages of using the least squares method

According to [9], if these four assumptions are not fulfilled, it can not only worsen the quality of the model used, but there is a possibility that the model will not be suitable for its practical use.

Mathematics is like a mill that grinds everything that is loaded into it . You can fill many pages with formulas, but if you proceed from erroneous assumptions, the final result will also contain erroneous conclusions.

According to [10], the least squares method is unstable to outliers. Estimates using the least squares method will coincide with estimates using the method of greatest similarity only if there is a nominal distribution of errors.

According to [11], the advantages and disadvantages of the least squares method are ( Fig. 1) .

**Methods**

The Fermat-Torricelli point is a point located from all the vertices of the triangle , so that the total distance from this point to each of the vertices of the triangle is minimal.  The task of finding Fermat-Torricelli points is used in econometrics, for example, to solve the problem of connecting three cities with roads of minimal total length. In the framework of this study, the author uses finding Fermat-Torricelli points to smooth numerical series.

Suppose we have a triangle ABC, then the point F will be the Fermat-Torricelli point if the condition is met

$$|AF| + |BF| + |CF| \to min \qquad (1)$$

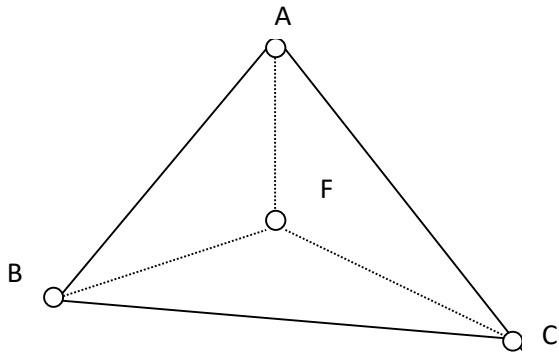

Fig. 2  The Fermat-Torricelli point

Theorem [12] Suppose a triangle has all angles less than $\frac{2\pi}{3}$ then the Fermat – Torricelli point has coordinates:

$x = \frac{X}{2d\sqrt{3}}$ , $y = \frac{Y}{2d\sqrt{3}}$ , where $d = \frac{r_{12}^2 + r_{13}^2 + r_{23}^2}{2} + |S| * \sqrt{3}$

$X = \sqrt{3}(x_1 r_{23}^2 + x_2 r_{13}^2 + x_3 r_{12}^2) + (x_1 + x_2 + x_3)|S| + 3sign(S)[(y_2 - y_1)(x_1 x_2 + y_1 y_2) +$

$+(y_1 - y_3)(x_1 x_3 + y_1 y_3) + (y_1 - y_3)(x_1 x_3 + y_1) + (y_3 - y_2)(x_2 x_3 + y_2 y_3)]$

$Y = \sqrt{3}(y_1 r_{23}^2 + y_2 r_{13}^2 + y_3 r_{12}^2) + (y_1 + y_2 + y_3)|S| + 3sign(S)[(x_2 - x_1)(x_1 x_2 + y_1 y_2) +$

$+(x_1 - x_3)(x_1 x_3 + y_1 y_3) + (x_1 - x_3)(x_1 x_3 + y_1) + (y_3 - y_2)(x_2 x_3 + y_2 y_3)]$

$S = x_1 y_2 + x_3 y_1 + x_2 y_3 - x_1 y_3 - x_2 y_1 - x_3 y_2$

$r_1, r_2, r_3$ — the length of the sides of the triangle, $x_1, x_2, x_3, y_1, y_2, y_3$ – coordinates of the vertices of the triangle. Suppose there is a numerical series consisting of eight points $x_1, x_2, x_3, x_4, x_5, x_6, x_7, x_8$ and we need to smooth it by constructing Fermat-Torricelli points, and then interpolate the Fermat-Torricelli points of some curve $f(x)$ (Fig. 3).

First, we build triangles from the points displaying the numerical series $x_1, x_2, x_3$, $x_2, x_3, x_4$, $x_3, x_4, x_5$, $x_4, x_5, x_6$, $x_5, x_6, x_7$, $x_6, x_7, x_8$. Then we find Fermat-Torricelli points for each of these triangles. These will be, respectively, points $f_1, f_2, f_3, f_4, f_5, f_6$.

Then, the curve is interpolated $f(x)$ by points $f_1, f_2, f_3, f_4, f_5, f_6$.

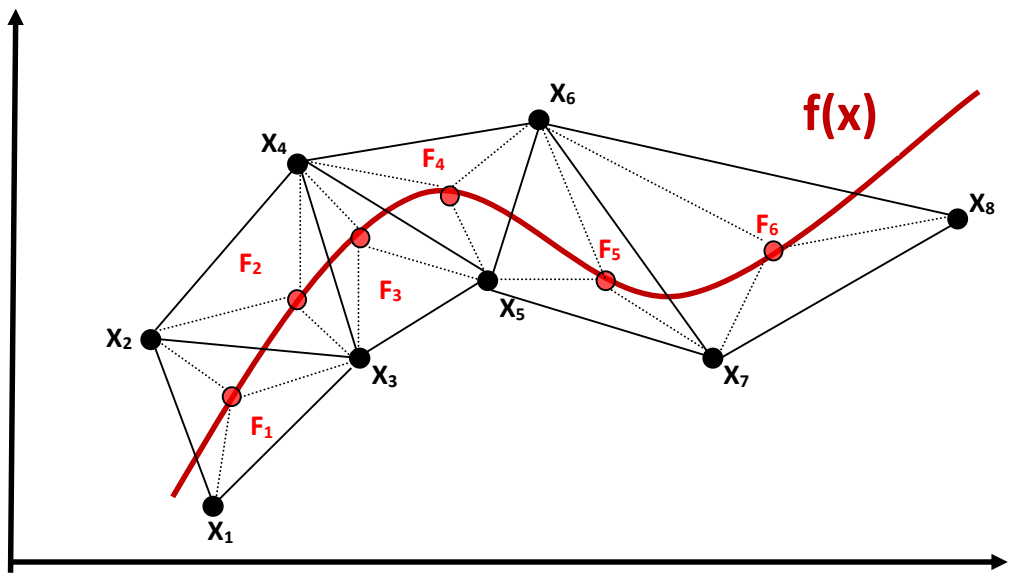

Fig.3 interpolating a curve $f(x)$ by Fermat-Torricelli points $f_1, f_2, f_3, f_4, f_5, f_6$.

To make calculations, we will use the inflation data in the Czech Republic in 2011-2021, which are available on [13] (Table 1)

Table.1 Inflation in the Czech Republic 2011 – 2021 [13]

| Year number | 1 | 2 | 3 | 4 | 5 | 6 |
|---|---|---|---|---|---|---|
| Year | 2011 | 2012 | 2013 | 2014 | 2015 | 2016 |
| Inflation, % | 2,2 | 3,5 | 1,4 | 0,4 | 0,3 | 0,6 |

| Year number | 7 | 8 | 9 | 10 | 11 |
|---|---|---|---|---|---|
| Year | 2017 | 2018 | 2019 | 2020 | 2021 |
| Inflation, % | 2,4 | 2 | 2,6 | 3,3 | 3,3 |

Using formula (2), we calculate the coordinates of Fermat-Torricelli points.

Table.2. Coordinates of Fermat-Torricelli points for the inflation chart in the Czech Republic 2011-2021

| Year number | 1,79128927 | 3 | 4 | 5 | 6 |
|---|---|---|---|---|---|
| Year | 2011,79128927 | 2013 | 2014 | 2015 | 2016 |
| Inflation, % | 2,46610159 | 1,4 | 0,4 | 0,3 | 0,6 |

| Year number | 7,12649666 | 8 | 9 | 10 | 10,6380343 |
|---|---|---|---|---|---|
| Year | 2017,12649666 | 2018 | 2019 | 2020 | 2020,6380343 |
| Inflation, % | 2,10452453 | 2 | 2,6 | 3,3 | 3,57193156 |

We show that the function

I =(0.264901377876643)*exp((0.249672956416996)*t)+(-0.007782663831297 + 0.015129431149835i)*exp((0.076090999247734 + 2.511250329378980i)*t)+(-0.007782663831297 - 0.015129431149835i)*exp((0.076090999247734 - 2.511250329378980i)*t)+(-1.671150941557596 - 0.869131660330525i)*exp((-0.303576461438207 + 1.138618581934044i)*t)+(-1.671150941557596 + 0.869131660330525i)*exp((-0.303576461438207 - 1.138618581934044i)*t)+(-0.014659833689592)*exp((-0.249672956416996)*t)+(0.138184785734736 - 0.180858361988375i)*exp((-0.076090999247734 - 2.511250329378980i)*t)+(0.138184785734736 + 0.180858361988375i)*exp((-0.076090999247734 + 2.511250329378980i)*t)+(0.003015325281862 - 0.001937822578385i)*exp((0.303576461438207 - 1.138618581934044i)*t)+(0.003015325281862 + 0.001937822578385i)*exp((0.303576461438207 + 1.138618581934044i)*t)          (2)

Interpolates a time series containing Fermat–Torricelli points for the inflation graph in the Czech Republic 2011-2021.

Really ,

| T (year number) | I (inflation) ( Tabl.2) | I (inflation) calculated by the formula (2) |
|---|---|---|
| 1,79128927 | 2,46610159 | 2,466101588 |
| 3 | 1,4 | 1,4 |
| 4 | 0,4 | 0,4 |
| 5 | 0,3 | 0,3 |
| 6 | 0,6 | 0,6 |
| 7,12649666 | 2,10452453 | 2,10452453 |
| 8 | 2 | 2 |
| 9 | 2,6 | 2,6 |
| 10 | 3,3 | 3,3 |
| 10,6380343 | 3,57193156 | 3,571931564 |

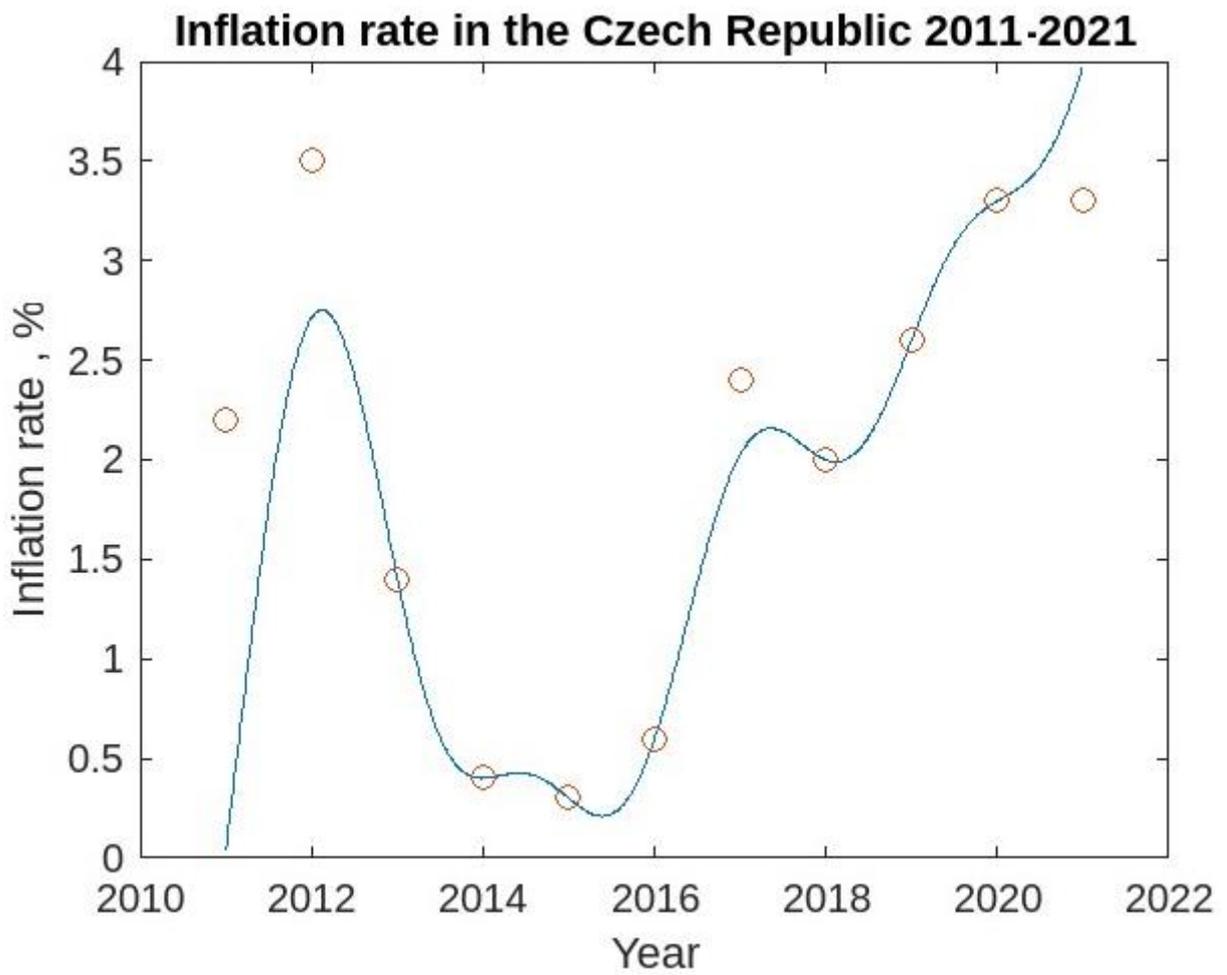

Fig. 3  Inflation rate in the Czech Republic 2010-2021

where    o - Statistical data on inflation in the Czech Republic for 2010-2021 [13]
,  ─────── - interpolation of Fermat-Torricelli points using the function  (2)

**Discussion**

    The use of regression analysis in the processing of experimental data is fraught with certain difficulties, which are caused by assumptions, namely the presence of a normal law of error distribution and statistical independence of variables.

    In practice, these conditions are not always met. And this may become prerequisites for the fact that the constructed model will not have practical value. As an alternative approach to the study of numerical series, according to the author, smoothing of numerical series using Fermat-Torricelli points with long-range interpolation of these points by series of exponents could be used.

## Conclusions

The use of exponential series for interpolation of numerical series makes it possible to achieve the accuracy of model construction no worse than regression analysis . At the same time, interpolation by series of exponents does not require that the errors of the numerical series obey the normal distribution law, as well as the statistical independence of variables. Interpolation of numerical series by exponential series represents a "black box" type model, that is, only input parameters and output parameters matter.